\documentclass[11pt,a4paper]{article}
\usepackage{amsmath,amssymb,amsfonts}
\usepackage{amsthm}
\usepackage{algorithmic}
\usepackage{graphicx}
\usepackage{textcomp}
\usepackage{xcolor}
\usepackage{authblk}

\newtheorem{definition}{Definition}
\newtheorem{theorem}{Theorem}
\newtheorem{property}{Property}
\newtheorem{constraint}{Constraint}

\usepackage{geometry}
\usepackage{setspace}
\usepackage{titlesec}
\titleformat{\section}{\normalfont\Large\bfseries}{\thesection}{1em}{}
\titleformat{\subsection}{\normalfont\large\bfseries}{\thesubsection}{1em}{}
\titleformat{\subsubsection}{\normalfont\normalsize\bfseries}{\thesubsubsection}{1em}{}

\title{A Proposal for High-Level Architectural Model Capable of \\
Expressing Various Data Collaboration Platform and Data Space Concepts}

\author[1]{Masaru Dobashi \texttt{Masaru.Dobashi@nttdata.com}} 
\author[1]{Kohei Toshimitsu \texttt{Kohei.Toshimitsu@nttdata.com}} 
\author[2]{Hirotsugu Seike \texttt{hirotsugu.seike@koshizuka-lab.org}} 
\author[1]{Miki Kanno \texttt{Miki.kanno@nttdata.com}} 
\author[2]{Genki Horie \texttt{genki.horie@koshizuka-lab.org}} 
\author[2]{Noboru Koshizuka \texttt{noboru@koshizuka-lab.org}}

\affil[1]{NTT DATA Group Corporation}
\affil[2]{University of Tokyo}

\date{}

\begin{document}

\maketitle

\begin{abstract}
This paper proposes "Data Space High-Level Architecture Model" (DS-HLAM) for 
expressing diverse data collaboration platforms across regional implementations. 
The framework introduces mathematically rigorous definitions with success conditions 
formalized through finite state automata theory, enabling interoperability while 
preserving digital sovereignty requirements.
\end{abstract}

\textbf{Keywords:} Data Spaces, High-Level Architecture, Data Collaboration Platform, Digital Sovereignty, Interoperability

\section{Overview}

Data Spaces represent technological and social mechanisms that provide services 
for exchanging and sharing data among multiple organizations and individuals 
in a trusted manner based on data sovereignty concepts~\cite{wef2025digital,giess2025discovering}. Regional Data Space 
initiatives have developed distinct architectural approaches shaped by local 
regulatory requirements and strategic priorities~\cite{ec2020data,cao2016basic,frontiers2025politics}. This diversity creates 
significant interoperability challenges across ecosystems~\cite{idsa2025papers}.

We propose a high-level architectural model called "Data Space High-Level 
Architecture Model" (DS-HLAM) that provides sufficient abstraction to 
accommodate diverse data collaboration approaches~\cite{cao2025society,mdpi2021society,idsa2024japan,dsa2025official} while maintaining essential 
functional elements required for inter-organizational data exchange and 
sharing~\cite{ieee2023dataex,meti2023ouranos}. The model captures fundamental relationships and interactions in 
data collaboration scenarios without prescribing specific implementation details.

\textbf{Note}: Detailed background on regional Data Space initiatives and related technologies 
will be provided in a separate document.

\section{Mathematical Definition}

\subsection{Data Space}
In DS-HLAM, Data Space is defined as the composite mechanism that consists of three elements as follows:

\begin{align}
DS &= \{Components, Data Service Methods, Data Governance Methods\} && \text{(Data Space)} \label{eq:dataspace}
\end{align}

\begin{itemize}
\item $Components$ represent subjects, objects, and medium for the Data Space 
such as organizations, data provision mechanisms, Data Units, and Social Mechanisms as described in the following subsection.

\item 
$Data Service Methods$ represent a set of operations to Data Spaces which provide data exchange and sharing.

\item 
$Data Governance Methods$ represent a set of operations to Data Spaces which enable enforcement of data governance in data exchange and sharing by controlling technical and social rules.
\end{itemize}

\begin{figure}[t]
\centering
\includegraphics[width=90mm]{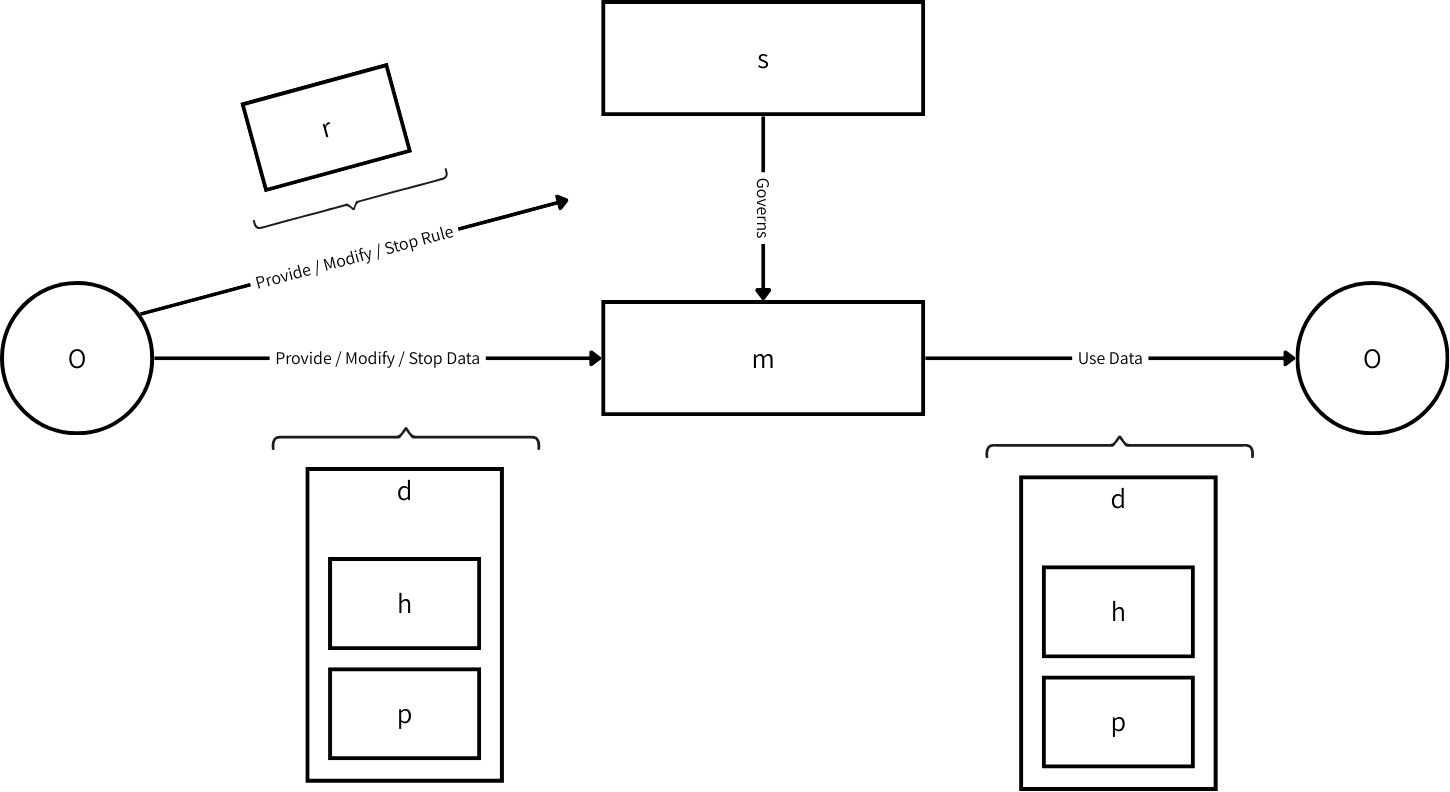}
\caption{Core Components and Interaction Patterns in DS-HLAM}
\label{DS-HLAM-Conceptual}
\end{figure}

Figure~\ref{DS-HLAM-Conceptual} illustrates the core components and interaction patterns of DS-HLAM, showing how Organizations exchange data through Data Service Methods and Data Governance Methods, with Social Mechanisms providing the underlying trust infrastructure.

\subsection{Components}
In DS-HLAM, Data Space contains five component sets as follows:

The Components are defined as follows:
\begin{align}
O &= \{o_1, o_2, \ldots, o_k\} && \text{(Organizations)} \label{eq:org_set} \\
M &= \{m_1, m_2, \ldots, m_l\} && \text{(Data Provision Mechanisms)} \label{eq:mech_set} \\
D &= \{d_1, d_2, \ldots, d_m\} && \text{(Data Units)} \label{eq:data_set} \\
S &= \{s_1, s_2, \ldots, s_n\} && \text{(Social Mechanisms)} \label{eq:social_set} \\
R &= \{r_1, r_2, \ldots, r_p\} && \text{(Rules)} \label{eq:rule_set}
\end{align}

In the many usual implementation, $d \in D$ consists of Header $h \in H$ and Payload $p \in P$ as follows:
\begin{align}
H &= \{h_1, h_2, \ldots, h_p\} && \text{(Headers)} \label{eq:header_set} \\
P &= \{p_1, p_2, \ldots, p_q\} && \text{(Payloads)} \label{eq:payload_set}
\end{align}
where each data unit $d_i \in D$ is composed as $d_i = (h_j, p_k)$ with $h_j \in H, p_k \in P$.

\subsubsection{Organizations}
Organizations ($O$) represent the participating entities in the data collaboration 
ecosystem, including enterprises, government agencies, academic institutions, 
and other stakeholders engaged in data sharing activities. 
They may have roles such as data providers, data users, authenticators, trust organizations, clearing houses, and data marketplaces.
In DS-HLAM, for model simplification, individuals are also defined as organizations (one-person organizations).

\subsubsection{Data Provision Mechanisms}
Data Provision Mechanisms ($M$) represent the technical and procedural methods which provide Data Units via Operations.
In usual Data Spaces, Data Provision Mechanisms may be implemented 
using database management systems, file systems, API endpoints for real-time streaming data, and so on. 
A data unit can be associated with one or more Data Provision Mechanisms.

\subsubsection{Data Units}
Data Units ($D$) represent the structured data assets being shared or exchanged,
which work as the fundamental units of value in data collaboration scenarios.

\begin{definition}[Header Structure]
\label{def:header_structure}
Each header $h \in H$ is defined as a logical object containing a set of typed attributes:
$$h = \{(t_1, v_1), (t_2, v_2), \ldots, (t_n, v_n)\}$$
where $(t_i, v_i) \in \text{AttributeType} \times \text{AttributeValue}$ 
represents control and governance information that is logically separate 
from payload content. The mandatory attribute type for Data Space integration is:
\begin{align}
t_{\text{social}} &\in \text{SocialMechanismType} \label{eq:social_attr}
\end{align}
Optional attribute types may include, but are not limited to:
\begin{align}
t_{\text{timestamp}} &\in \text{TimestampType} \label{eq:timestamp_attr} \\
t_{\text{format}} &\in \text{DataFormatType} \label{eq:format_attr}
\end{align}
\end{definition}

Additional optional attributes can encompass access control specifications, 
quality metrics, provenance information, lifecycle management parameters, 
and domain-specific metadata requirements depending on the particular data 
collaboration use case and implementation context. The mandatory social 
mechanism attribute reflects the fundamental requirement for Data Spaces 
and modern data collaboration platforms to operate within appropriate 
governance and trust frameworks.

Headers ($H$) contain metadata for the following two objectives. 
First, they are used for finding data units in a data space by organizations.
Second, they are used for control metadata including authentication tokens, access control specifications, regulatory compliance attestations, and traceability chain identifiers that support data sovereignty by enabling organizations to control data disclosure and access permissions across organizational boundaries. 
Unlike network protocol headers that dictate physical packet structure, 
headers in this model serve as logical objects that carry decision-making criteria, policy specifications, and coordination information that work in conjunction with social mechanisms to determine the feasibility and authorization of data sharing activities. 
This abstraction allows headers to be implemented across diverse technological platforms—from basic file metadata and HTTP headers to API specifications, database schemas, message queue attributes, and advanced implementations including blockchain smart contract parameters—while maintaining consistent semantic meaning for data collaboration governance. 

% REVISION: [2025-08-10] 非構造データへの言及を追加
Payloads ($P$) encompass both structured and unstructured data content. Structured payloads include business data such as product specifications, manufacturing parameters, supply chain information, and carbon footprint measurements as specified in data collaboration guidelines for supply chain traceability~\cite{ipa2023guideline,ipa2024guideline}. Unstructured payloads incorporate diverse content types including documents, images, videos, sensor streams, and other binary data formats that may require specialized processing or interpretation mechanisms. This comprehensive definition ensures the model can represent the full spectrum of data types exchanged in modern data collaboration scenarios, from traditional database records to multimedia content and IoT sensor data.

\subsubsection{Social Mechanisms}

Social Mechanisms ($S$) represent the technological and social infrastructure that provides the foundational capabilities for realizing data governance.
They encompass trust relationships, governance frameworks, and collaborative agreements that enable data sharing, including digital trust services such as European Trust Services under eIDAS Regulation, Japan's GBizID\footnote{As of June 2025, Japan's GBizID is under discussion for expansion to private sector usage, as outlined in the Digital Agency's policy framework for digital identity infrastructure~\cite{digital2025gbizid}.}, and other verification and assurance services that validate identities, credentials, and authorization for data collaboration activities.

At the highest level of abstraction, social mechanisms provide the foundational infrastructure that supports multiple overlapping functional dimensions, including identity verification and authentication capabilities, policy enforcement and access control mechanisms, and legal compliance and certification requirements. As a component set, Social Mechanisms provide the stable technological and social foundation upon which Rules ($R$) operate. The specific categorization and decomposition of social mechanisms depend on the particular data collaboration context, regulatory environment, and implementation requirements.

\subsubsection{Rules}

Rules ($R$) represent the dynamic governance logic and operational policies that govern data collaboration activities within Data Space. Unlike Social Mechanisms which provide the foundational infrastructure, Rules are configurable and modifiable elements that define specific conditions, constraints, and actions for data sharing operations. Rules encompass access control policies, data usage constraints, transformation requirements, compliance conditions, and business logic that can be established, modified, or terminated through governance operations. 

While Rules primarily operate within the Social Mechanisms infrastructure, they can also interact with Data Provision Mechanisms when technical-level enforcement is beneficial. This optional coupling allows Rules to be enforced at the most appropriate architectural layer—whether at the governance level for policy decisions or at the mechanism level for operational efficiency. The separation of Rules from Social Mechanisms enables flexible governance adaptation while maintaining stable infrastructure foundations.

\subsubsection{Relationship with Operations}

The five component sets work together to enable the execution of Data Service Methods and Data Governance Methods defined in the following sections. Operations utilize these components in the following manner:

\begin{itemize}
\item \textbf{Data Service Methods} utilize multiple components simultaneously: Data Provision Mechanisms (M) provide the technical infrastructure for data handling, Organizations (O) define the actors and their roles, Data Units (D) represent the information being exchanged, Social Mechanisms (S) provide the governance infrastructure, and Rules (R) define the specific policies and constraints.

\item \textbf{Data Governance Methods} operate on Rules (R) that execute within the Social Mechanisms (S) infrastructure, while considering organizational responsibilities (O), data context (D), and technical constraints imposed by Data Provision Mechanisms (M). The Social Mechanisms provide the foundational platform, while Rules (R) represent the dynamic governance logic that can be established, modified, or terminated through governance operations.
\end{itemize}

This collaborative utilization of components ensures that operations maintain both technical feasibility and governance compliance throughout data collaboration processes.

\subsection{Data Service Methods}
In DS-HLAM, Data Space contains Data Service Methods as follows:

\subsubsection{Provide Data}

Provide Data ($Provide\_Data()$) is a method which requests to put a new data unit $d$ to a data space by organization $o$
in the condition designated by $cond$.
If this request succeeds, 
return value $ret$ is $1$, and a new data unit $d$ is added to the set of data units $D$, 
then the new set of data units is expressed as $D'$.
If it fails, return value $ret$ is $0$, and no operation is performed.
A parameter $cond$ means some parameter representing additional conditions when this method is called.

\begin{align}
ret = Provide\_Data (o, d, cond): D \rightarrow D' && \text{where } D' = 
    \begin{cases}
    D \cup \{d\}  & (ret = 1) \\
    D  \text{ (unchanged)} & (ret = 0)
    \end{cases} \label{eq:provide_data} 
\end{align}

\subsubsection{Modify Data}

Modify Data ($Modify\_Data()$) is a method which requests to modify the data provision service conditions for a data unit $d$ in a data space by organization $o$
in the condition designated by $cond$.
This represents changes to access permissions, usage policies, or provision mechanisms rather than alterations to the data unit content itself.
If this request succeeds, 
return value $ret$ is $1$, and the data unit $d$ in the set of data units $D$ is updated with modified provision conditions, 
then the new set of data units is expressed as $D'$.
If it fails, return value $ret$ is $0$, and no operation is performed.

\begin{align}
ret = Modify\_Data (o, d, d', cond): D \rightarrow D' && \text{where } D' = 
    \begin{cases}
    (D \setminus \{d\}) \cup \{d'\}  & (ret = 1) \\
    D  \text{ (unchanged)} & (ret = 0)
    \end{cases} \label{eq:modify_data} 
\end{align}

\subsubsection{Stop Data}

Stop Data ($Stop\_Data()$) is a method which requests to terminate the data provision service for a data unit $d$ in a data space by organization $o$
in the condition designated by $cond$.
This operation stops the data sharing service while maintaining data sovereignty principles, allowing organizations to control when and how their data is no longer accessible to other participants.
If this request succeeds, 
return value $ret$ is $1$, and the data provision service for data unit $d$ is terminated, effectively removing it from the accessible set of data units $D$, 
then the new set of data units is expressed as $D'$.
If it fails, return value $ret$ is $0$, and no operation is performed.

\begin{align}
ret = Stop\_Data (o, d, cond): D \rightarrow D' && \text{where } D' = 
    \begin{cases}
    D \setminus \{d\}  & (ret = 1) \\
    D  \text{ (unchanged)} & (ret = 0)
    \end{cases} \label{eq:stop_data} 
\end{align}

\subsubsection{Use Data}

Use Data ($Use\_Data()$) is a method which requests to use a data unit $d$ in a data space by organization $o$
in the condition designated by $cond$.
If this request succeeds, 
return value $ret$ is $1$, and the organization $o$ uses the data unit $d$ in the data space, 
and there is no side effect to the data space.
If it fails, return value $ret$ is $0$, and no operation is performed.

\begin{align}
ret = Use\_Data (o, d, cond): D \rightarrow D && \text{where } ret \in \{0, 1\} \label{eq:use_data} 
\end{align}

\subsection{Data Governance Methods}
In DS-HLAM, Data Governance Methods operate on a set of governance rules $R = \{r_1, r_2, \ldots, r_n\}$ that coordinate with the Social Mechanisms infrastructure. These rules represent the dynamic governance logic, while Social Mechanisms provide the stable technological and social foundation that enables rule coordination and execution.

Data Space contains three Data Governance Methods as follows:

\subsubsection{Provide Rule}

Provide Rule ($Provide\_Rule()$) is a method which requests to establish a new governance rule $r$ in a data space by organization $o$
in the condition designated by $cond$.
If this request succeeds, 
return value $ret$ is $1$, and a new rule $r$ is added to the set of governance rules $R$ that coordinate with the Social Mechanisms infrastructure, 
then the new set of governance rules is expressed as $R'$.
If it fails, return value $ret$ is $0$, and no operation is performed.

\begin{align}
ret = Provide\_Rule (o, r, cond): R \rightarrow R' && \text{where } R' = 
    \begin{cases}
    R \cup \{r\}  & (ret = 1) \\
    R  \text{ (unchanged)} & (ret = 0)
    \end{cases} \label{eq:provide_rule} 
\end{align}

\subsubsection{Modify Rule}

Modify Rule ($Modify\_Rule()$) is a method which requests to replace a governance rule $r$ with a new rule $r'$ in a data space by organization $o$
in the condition designated by $cond$.
If this request succeeds, 
return value $ret$ is $1$, and the rule $r$ in the set of governance rules $R$ is replaced with a new rule $r'$, 
then the new set of governance rules is expressed as $R'$.
If it fails, return value $ret$ is $0$, and no operation is performed.

\begin{align}
ret = Modify\_Rule (o, r, r', cond): R \rightarrow R' && \text{where } R' = 
    \begin{cases}
    (R \setminus \{r\}) \cup \{r'\}  & (ret = 1) \\
    R  \text{ (unchanged)} & (ret = 0)
    \end{cases} \label{eq:modify_rule} 
\end{align}

\subsubsection{Stop Rule}

Stop Rule ($Stop\_Rule()$) is a method which requests to terminate a governance rule $r$ in a data space by organization $o$
in the condition designated by $cond$.
If this request succeeds, 
return value $ret$ is $1$, and the rule $r$ in the set of governance rules $R$ is terminated, 
then the new set of governance rules is expressed as $R'$.
If it fails, return value $ret$ is $0$, and no operation is performed.

\begin{align}
ret = Stop\_Rule (o, r, cond): R \rightarrow R' && \text{where } R' = 
    \begin{cases}
    R \setminus \{r\}  & (ret = 1) \\
    R  \text{ (unchanged)} & (ret = 0)
    \end{cases} \label{eq:stop_rule} 
\end{align}

Here, $ret \in \{0,1\}$ represents whether the governance method operation is successfully executed or not.
The success of these operations typically depends on the coordination between the governance rules $R$ and the Social Mechanisms infrastructure $S$, as well as organizational permissions within a data space.

\subsection{Transaction Model}

Building upon the Data Service Methods and Data Governance Methods defined in the previous sections, we now formalize how the success and failure combinations of these individual operations determine the overall success of data collaboration transactions between organizations.

\begin{definition}[Data Collaboration Transaction]
\label{def:transaction}
A data collaboration transaction represents a sequence of operations from the seven fundamental methods defined above. The success of the overall transaction depends on the appropriate combination and sequencing of these operations:
\begin{align}
T_{success} &= f(Op_1, Op_2, \ldots, Op_n) \nonumber \\
\text{where } Op_i &\in \{Provide\_Data, Modify\_Data, Stop\_Data, Use\_Data, \nonumber \\
&\quad\quad\quad Provide\_Rule, Modify\_Rule, Stop\_Rule\} \label{eq:transaction_operations}
\end{align}
\end{definition}

The fundamental principle is that successful data collaboration emerges from the coordinated execution of these concrete operations, rather than from abstract success conditions.

\subsubsection{Basic Data Exchange Pattern}

The most fundamental data collaboration pattern involves the successful execution of data provision and consumption operations:

\begin{theorem}[Basic Data Exchange Success Condition]
\label{thm:basic_success_condition}
A basic data exchange transaction between organizations $o_p$ (provider) and $o_u$ (user) for data unit $d$ is successful if and only if:
\begin{equation}
\begin{split}
\text{Success}_{basic}(o_p, o_u, d) = &Provide\_Data(o_p, d, cond_p) = 1 \\
&\land Use\_Data(o_u, d, cond_u) = 1 \\
&\land \text{Compatible}(cond_p, cond_u) = 1
\end{split}
\label{eq:basic_success_condition}
\end{equation}
\end{theorem}

\noindent where $Compatible(cond_p, cond_u)$ represents the compatibility between the provision conditions set by the provider organization and the usage conditions required by the user organization.

\subsubsection{Evolving Governance Transactions}

More sophisticated data collaboration scenarios involve governance operations that establish, modify, or terminate rules as the data exchange progresses:

\begin{definition}[Evolving Governance Transaction Pattern]
\label{def:governance_transaction}
An evolving governance transaction involves the coordinated execution of data service operations alongside governance operations that adapt rules throughout the collaboration process:
\begin{align}
\text{Success}_{governance}(T) &= \text{Success}_{basic}(o_p, o_u, d) \nonumber \\
&\land \bigwedge_{i} GovernanceOp_i = 1 \label{eq:governance_success}
\end{align}

\noindent where $GovernanceOp_i \in \{Provide\_Rule, Modify\_Rule, Stop\_Rule\}$ represents the governance operations required for the specific collaboration context.
\end{definition}

This pattern captures the reality that governance frameworks must evolve throughout the collaboration lifecycle, adapting to changing requirements, regulatory updates, or organizational policies while maintaining data sovereignty principles.

\subsubsection{Dynamic Transaction Management}

The transaction model also accommodates dynamic management scenarios where ongoing data collaborations require modification or termination:

\begin{definition}[Dynamic Management Operations]
\label{def:dynamic_management}
For established data collaborations, dynamic management operations enable runtime adjustments:
\begin{align}
\text{Success}_{modify}(o, d) &= Modify\_Data(o, d, cond_{new}) = 1 \label{eq:modify_success} \\
\text{Success}_{stop}(o, d) &= Stop\_Data(o, d, cond_{stop}) = 1 \label{eq:stop_success}
\end{align}
subject to the constraint that only data-providing organizations can successfully execute these operations.
\end{definition}

This comprehensive transaction model demonstrates how the success of complex data collaboration scenarios emerges from the coordinated execution of the seven fundamental operations, providing a concrete foundation for implementing and reasoning about data sovereignty in distributed collaboration environments.

\subsubsection{Operational Transaction Lifecycle}

To provide a concrete operational perspective on how the seven fundamental operations coordinate to achieve successful data collaboration, we model the transaction lifecycle as a sequence of state transitions driven by the execution of specific operations.

% NOTE: データスペース全体ではなく個別取引の状態遷移を表現
% q_2 (Validated) は「データのアテステーション」ではなく「取引全体の検証」
% Modifyはデータ自体ではなく提供条件/メタデータの更新
\begin{definition}[Transaction Lifecycle Automaton]
\label{def:transaction_automaton}
A transaction lifecycle automaton represents the execution sequence of the seven fundamental operations. The automaton states correspond to the operational status of data collaboration:
\begin{itemize}
\item $q_0$: Initial (no data provision active)
\item $q_1$: Data Provided (Provide\_Data executed successfully)
\item $q_2$: Rules Established (governance rules in place)
\item $q_f$: Active Collaboration (Use\_Data successful, collaboration ongoing)
\item $q_{mod}$: Under Modification (Modify operations in progress)
\item $q_{stop}$: Terminated (Stop operations executed)
\end{itemize}
\end{definition}

The key insight is that successful data collaboration emerges from the proper sequencing of concrete operations:

\begin{align}
q_0 \xrightarrow{Provide\_Data(o_s, d, cond_s) = 1} &q_1 \\
q_1 \xrightarrow{Provide\_Rule(o, r, cond_r) = 1} &q_2 \\
q_2 \xrightarrow{Use\_Data(o_c, d, cond_c) = 1} &q_f \\
q_f \xrightarrow{Modify\_Data(o_s, d, cond_{mod}) = 1} &q_{mod} \\
q_{mod} \xrightarrow{Use\_Data(o_c, d, cond_c) = 1} &q_f \\
q_* \xrightarrow{Stop\_Data(o_s, d, cond_{stop}) = 1} &q_{stop} \\
q_{mod} \xrightarrow{Stop\_Rule(o_s, r, cond_{stop}) = 1} &q_{stop}
\end{align}

This operational model demonstrates how the abstract concept of "transaction success" is grounded in the concrete execution of the seven fundamental operations defined in the previous sections.

\begin{theorem}[Operational Success Condition]
\label{thm:operational_success}
A data collaboration is successful if and only if there exists a valid sequence of operations from the seven fundamental methods that leads from the initial state $q_0$ to the active collaboration state $q_f$:
$$\text{Success}(collaboration) \Leftrightarrow \exists \text{ valid operation sequence: } q_0 \rightarrow^* q_f$$
\end{theorem}

The transaction lifecycle automaton involves key states: initial state ($q_0$), data provided ($q_1$), rules established ($q_2$), active collaboration ($q_f$), under modification ($q_{mod}$), and terminated ($q_{stop}$). Critical transitions include: after modifications ($q_{mod} \rightarrow q_f$), data can be used again via $Use\_Data()$ operations, while termination from modification state ($q_{mod} \rightarrow q_{stop}$) can occur through both $Stop\_Data()$ and $Stop\_Rule()$ operations.

This operational perspective emphasizes that data collaboration success is not determined by abstract properties, but by the concrete execution of specific operations. The model provides several practical advantages:

\begin{itemize}
\item \textbf{Implementation Guidance}: Clear operational steps for achieving data collaboration
\item \textbf{Debugging Support}: Failed collaborations can be traced to specific operation failures
\item \textbf{Sovereignty Preservation}: Each operation explicitly respects organizational control
\item \textbf{Scalability}: Operations can be composed and sequenced for complex scenarios
\end{itemize}

\subsubsection{Operational Constraints}

To ensure data sovereignty and logical consistency, the execution of the seven fundamental operations must satisfy the following constraints:

\begin{constraint}[Operation Precedence Constraints]
\label{constraint:precedence}
The following precedence relationships must hold for valid operation sequences:
\begin{align}
Use\_Data(o_c, d, cond) = 1 &\Rightarrow \exists o_s : Provide\_Data(o_s, d, cond_s) = 1 \label{eq:use_precedence} \\
Modify\_Data(o, d, cond) = 1 &\Rightarrow Provide\_Data(o, d, cond_{orig}) = 1 \label{eq:modify_precedence} \\
Stop\_Data(o, d, cond) = 1 &\Rightarrow Provide\_Data(o, d, cond_{orig}) = 1 \label{eq:stop_precedence}
\end{align}
\end{constraint}

These constraints ensure: (1) \textbf{Data Sovereignty}: Only data-providing organizations can modify or stop their data provision services, (2) \textbf{Logical Consistency}: Data cannot be used unless it has been provided, and (3) \textbf{Operational Integrity}: The system maintains consistent state throughout operation sequences.

To illustrate the practical applicability of our proposed model, we provide preliminary instantiation examples of three major Data Space implementations: the Ouranos Ecosystem's Battery Traceability Platform, IDS RAM with the Dataspace Protocol, and the Ouranos-Catena-X interoperability experiment. These initial case studies suggest the model's potential to represent diverse architectural approaches while maintaining consistent abstraction levels. Detailed instantiation examples and exploratory analyses will be documented separately, though comprehensive validation remains future work.

\section{How to Use the Model}

\subsection{Refinement Process (Vertical Consistency)}

The model employs a functional abstraction approach that separates conceptual relationships from implementation-specific details, enabling diverse data collaboration platforms to be represented within the same framework while accommodating different technological choices, regulatory requirements, and organizational practices.

A critical feature of this architectural approach is its support for hierarchical decomposition and progressive refinement. Each element within the fundamental sets can contain components and subsystems that enable stepwise elaboration from high-level architectural concepts to implementation-specific details. This hierarchical capability allows the model to capture initiative-specific and use case-specific characteristics through progressive decomposition while maintaining architectural consistency across abstraction levels.

\begin{property}[Consistency Across Abstraction Levels]
\label{prop:consistency}
The consistency property addresses two distinct but related scenarios for maintaining success conditions across different contexts.
\end{property}

\subsubsection{Constraint-Preserving Decomposition}

To formalize the constraint-preserving property within individual frameworks, we define the mathematical foundation for hierarchical decomposition that enables systematic refinement from abstract specifications to concrete implementations.

\begin{definition}[Hierarchical Decomposition]
\label{def:hierarchical_decomposition}
For any set $X \in \{O, M, D, S\}$, there exists a decomposition function:
$$\text{Decompose}: X \rightarrow \mathcal{P}(X)$$
such that for any element $x \in X$:
$$x = \bigcup_{x_i \in \text{Decompose}(x)} x_i$$
where $\mathcal{P}(X)$ denotes the power set of $X$, and $\text{Decompose}(x) \neq \emptyset$ implies progressive refinement from high-level concepts to implementation-specific details.
\end{definition}

The preservation of success conditions across abstraction levels is not automatic but depends on maintaining functional equivalence during decomposition. For any decomposition $\{x_1, x_2, \ldots, x_k\} = \text{Decompose}(x)$, success condition preservation requires constraint-preserving decomposition:
\begin{equation}
\begin{split}
\text{Success}(T) = \text{Success}(T') \text{ iff } \\
\text{ConstraintPreserving}(\text{Decompose}(d,m,s))
\end{split}
\label{eq:appropriate_decomposition}
\end{equation}
where $T$ and $T'$ represent transactions at different abstraction levels with $T'$ containing decomposed components $d' \in \text{Decompose}(d)$, $m' \in \text{Decompose}(m)$, and $s' \in \text{Decompose}(s)$.

ConstraintPreserving decomposition maintains: (1) functional equivalence of capabilities, (2) preservation of security and regulatory constraints, and (3) consistency of operational context between abstraction levels.

\begin{figure}[t]
\centering
\includegraphics[width=90mm]{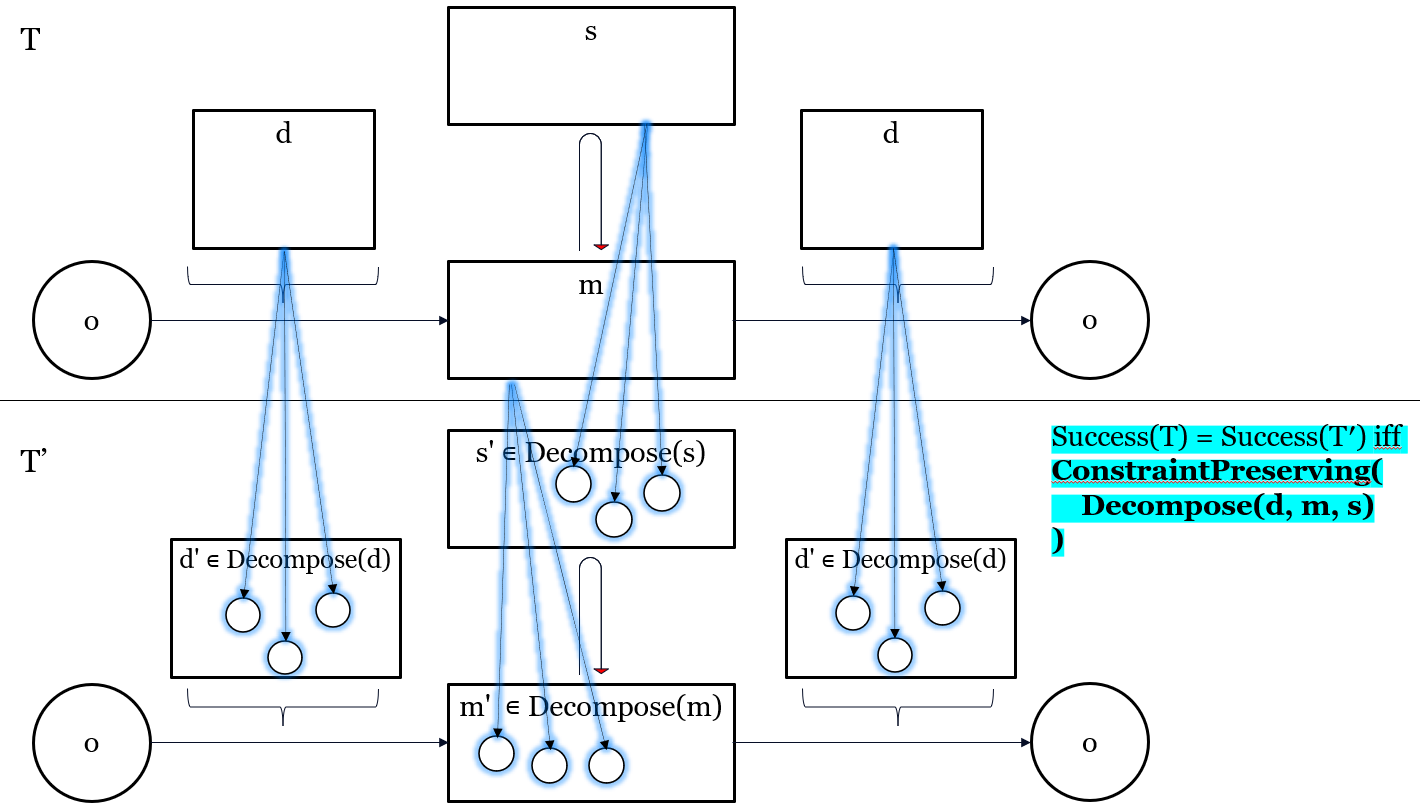}
\caption{ConstraintPreserving}
\label{ConstraintPreserving}
\end{figure}

\subsection{Interoperability Process (Horizontal Consistency)}

Beyond maintaining consistency within individual frameworks through constraint-preserving decomposition, achieving interoperability across heterogeneous regional implementations requires mutual recognition mechanisms.

For heterogeneous implementations across different regional or conceptual frameworks, mutual recognition enables cross-framework consistency through:
\begin{equation}
\text{MutualRecognition}(s_i, s_j) = 1 \Rightarrow \text{Success}(T_i) \leftrightarrow \text{Success}(T_j)
\label{eq:mutual_recognition}
\end{equation}
where $s_i$ and $s_j$ represent social mechanisms from different frameworks, enabling interoperability while preserving sovereignty requirements at each abstraction level.

\begin{figure}[t]
\centering
\includegraphics[width=90mm]{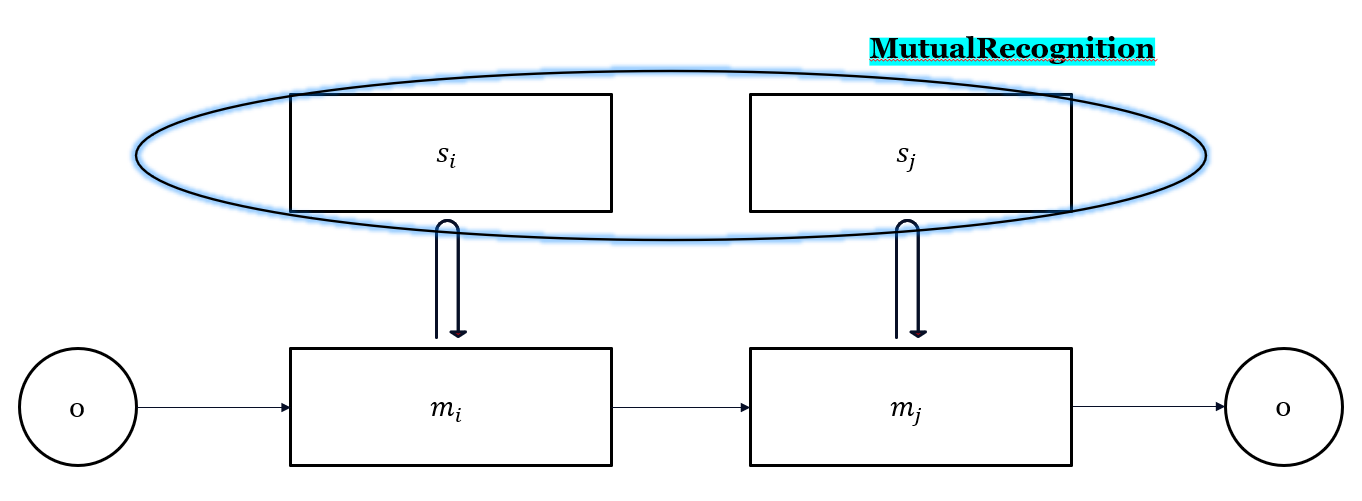}
\caption{MutualRecognition}
\label{MutualRecognition}
\end{figure}

\subsubsection{Interoperability of Data Exchange and Sharing Functions}

One important objective of this high-level architecture is to provide a foundation for interoperability among diverse data spaces. Here, we have two different Data Spaces: $DataSpace_A$ and $DataSpace_B$. 

A data unit $d_A$ is provided by $Provide\_Data_A()$, and there exists a set of parameters $\exists (o_i, cond_i)$ that satisfies the following condition:
$$Use\_Data_A(o_i, d_A, cond_i) = 1$$
(i.e., someone can use the data $d_A$).

In this situation, if there exists a set of parameters $\exists (o_j, cond_j)$ that satisfies the following condition:
$$Use\_Data_B(o_j, d_A, cond_j) = 1$$
then $DataSpace_A$ and $DataSpace_B$ have "Interoperability" of Data Exchange and Sharing Functions.

This means that when data is provided to $DataSpace_A$ and becomes usable, if there exists some way to use that data from $DataSpace_B$ (without directly providing it to $DataSpace_B$), then there is interoperability in data exchange and sharing between the two data spaces. The "some way to use" can be defined as the existence of parameters that make $Use\_Data(params) = 1$.

\textbf{Interoperability vs. Multi-use Scenarios}: When mutual recognition establishes interoperability between data spaces, users can access data through unified interfaces. Alternatively, multi-space usage patterns—where users independently interact with different data spaces using their respective protocols—represent a pragmatic approach that accommodates technological diversity and innovation. This flexibility allows superior specifications and technologies to emerge naturally while enabling users to choose optimal solutions for specific contexts. Both approaches serve different strategic objectives: interoperability for seamless integration, and multi-use for technological advancement and user choice flexibility. Regardless of the chosen approach, the scientific explicability of underlying concepts and specifications remains fundamental to ensuring reliable and trustworthy data collaboration systems.

\section{Conclusion}

% REVISION: [2025-07-21] 簡潔化・Future Work分離
This paper presented DS-HLAM (Data Space High-Level Architecture Model) with 
five fundamental component sets—Organizations (O), Data Provision Mechanisms (M), 
Data Units (D), Social Mechanisms (S), and Rules (R)—where Data Units are 
typically composed of Headers (H) and Payloads (P). The model provides 
mathematically rigorous success conditions for expressing diverse Data Space concepts. 
The model distinguishes between stable infrastructure (Social Mechanisms) and 
dynamic governance logic (Rules), enabling flexible policy adaptation. 
The framework includes both Data Service Methods (Provide, Modify, Stop, Use) 
and Data Governance Methods (Provide Rule, Modify Rule, Stop Rule) with 
mathematically defined success conditions. The model's hierarchical decomposition 
enables consistent representation across abstraction levels while accommodating 
regional implementation diversity through mutual recognition mechanisms.

Preliminary application through three exploratory case studies—the Ouranos 
Ecosystem's Battery Traceability Platform, IDS RAM with the Dataspace 
Protocol, and the Ouranos-Catena-X interoperability experiment—suggests 
the model's potential to unify heterogeneous data collaboration frameworks 
(detailed examples to be documented separately). These initial explorations indicate 
that diverse regional approaches—from Japan's Society 5.0 to Europe's 
Data Spaces—may be systematically represented within our framework, 
potentially providing a foundation for cross-regional interoperability.

The proposed model addresses critical challenges in current data collaboration 
architectures: fragmentation across regional implementations, insufficient 
abstraction levels, and the tension between sovereignty and interoperability. 
By enabling mutual recognition mechanisms while preserving regional autonomy, 
the framework contributes to realizing a global data economy that achieves 
both technological compatibility and socio-political sustainability.

\subsection{Future Work}

% REVISION: [2025-07-21] 元論文から具体的課題を抽出
Three critical research directions emerge from this work:

\textbf{Model Concretization}: Development of concrete instantiations for 
the Provide, Use, and Attestation functions with measurable criteria for 
specific domains will enable practical implementation and automated validation. 
This includes defining domain-specific capability assessments, compliance 
verification mechanisms, and trust quantification metrics.

\textbf{Empirical Validation}: Comprehensive validation of the proposed 
model through real-world implementations remains a critical priority. 
Leveraging ongoing Ouranos-Catena-X interoperability experiments~\cite{catena2024mou,nttdata2025battery,catena2025interop} and 
other operational deployments~\cite{ipa2025ouranos} will provide empirical evidence for model 
effectiveness. Systematic evaluation in diverse operational scenarios 
will yield insights for model refinement and implementation guidelines.

\textbf{Standardization Frameworks}: Development of standardized metrics 
for data collaboration effectiveness and architectural quality assessment~\cite{iso42010,iso42030,law2025iso} 
will enhance practical utility. This includes evaluation frameworks that 
address both collaborative outcomes and the architectural qualities 
enabling effective inter-organizational data sharing.

% Appendix temporarily removed for paper length adjustment
% \clearpage
% \input{appendix_case_studies}

\end{document}